\documentclass[%
twocolumn,
groupedaddress,
amsmath,amssymb,
aps,
pra,
]{revtex4-2}

\usepackage{graphicx}
\usepackage{amsfonts}
\usepackage{bm}
\usepackage{hyperref,url}
\usepackage{changes}
\usepackage{xcolor}
\usepackage{soul}

\newcommand{\bd}{\begin{displaymath}}
\newcommand{\ed}{\end{displaymath}}
\newcommand{\be}{\begin{equation}}
\newcommand{\ee}{\end{equation}}
\newcommand{\ba}{\begin{eqnarray}}
\newcommand{\ea}{\end{eqnarray}}

\begin{document}

\title[]{Decoherence in matter-wave Talbot interference:\\ A hydrodynamic probability-flow analysis}

\author{David Navia}
\affiliation{Department of Optics, Faculty of Physical Sciences,
Universidad Complutense de Madrid,\\
Pza.\ Ciencias 1, Ciudad Universitaria E-28040 Madrid, Spain}

\author{A. S. Sanz}
\affiliation{Department of Optics, Faculty of Physical Sciences,
Universidad Complutense de Madrid,\\
Pza.\ Ciencias 1, Ciudad Universitaria E-28040 Madrid, Spain}

\begin{abstract}
We investigate the suppression of matter-wave Talbot interference under environmentally induced decoherence. The system is modeled as an atomic beam diffracted by a periodic grating whose transverse dynamics is described within the paraxial approximation. Environmental coupling is introduced through an effective open-system model that exponentially damps spatial coherences between diffracted components, allowing a continuous interpolation between the coherent Talbot regime and the incoherent far-field diffraction limit.
In addition to the usual intensity and transverse-momentum distributions, we analyze the local probability flow associated with the diffracted matter wave. The corresponding Bohmian, or hydrodynamic, representation is used here as a diagnostic tool fully equivalent to the standard quantum description, with no additional assumptions beyond the probability current of the paraxial wave field.
In the present Talbot geometry, this analysis shows how decoherence progressively suppresses the carpet structure and smooths the transverse-momentum distribution, while the flow may remain organized into channels determined by the grating periodicity.
The results illustrate, in a periodic matter-wave Talbot geometry, that the loss of visible interference and the degradation of the grating-defined flow-domain organization need not occur simultaneously.
In particular, flux-channel structures can persist in parameter regimes where multislit interference features have already been strongly reduced. This distinction provides a local characterization of decoherence in matter-wave Talbot interferometry and complements previous trajectory-based analyses of coherence loss in simpler interference and confined geometries.
\end{abstract}




\maketitle


\section{Introduction}
\label{sec1}

Matter-wave interferometry provides one of the most sensitive probes of spatial coherence in atomic, molecular, and mesoscopic quantum systems.
Over the last few decades, diffraction and interferometric techniques with atoms and molecules have become central tools in atomic, molecular and optical (AMO) physics, enabling precision measurements, tests of fundamental quantum mechanics, and controlled studies of the quantum-to-classical transition \cite{cronin:RMP:2009,arndt:RMP:2012,Kialka:AVSQuantSci:2022}. In this context, near-field interferometers based on periodic gratings, including Talbot and Talbot--Lau configurations, are particularly relevant because of their robustness, scalability to increasingly massive particles, and high sensitivity to spatial coherence \cite{Hornberger:PRA:2004,arndt:NatCommun:2011,arndt:NNanotech:2012}.

The Talbot effect is a self-imaging phenomenon that arises when a coherent wave is diffracted by a periodic structure. Originally observed in classical optics \cite{talbot:PhilosMag:1836} and later explained in terms of Fresnel diffraction \cite{rayleigh:PhilosMag:1881}, it has since become a paradigmatic example of near-field interference, with close connections to wave-packet revivals and quantum carpets \cite{patorski:ProgOpt:1989,berry:JModOpt:1996,berry:PhysWorld:2001,xiao:AdvOptPhoton:2013}. In matter-wave optics, Talbot interference provides a spatially resolved manifestation of phase coherence across many diffraction orders and is therefore especially well suited to analyze how environmental interactions degrade interference in extended periodic systems \cite{cronin:RMP:2009,Hornberger:PRA:2004,schleich:NJP:2013}.

Decoherence in matter-wave interferometry has been extensively investigated both theoretically and experimentally. Environmental scattering, photon recoil, thermal radiation, dipole-dipole interactions, and long-range environmental couplings can reduce fringe visibility and drive the gradual emergence of classical-like behavior \cite{walls:PRA:1993,walls:Nature:1994,chapman:PRL:1995,hornberger:PRL:2003,Hornberger:PRA:2004,lombardo:PRA:2005,Kunjummen:PRA:2023,Fragolino:PRA:2024}. These mechanisms are especially relevant in modern interferometers involving high-mass matter waves, e.g., large molecules, clusters, and nanoparticles, where the competition between coherent diffraction and environmental localization becomes increasingly delicate \cite{arndt:Nature:1999,arndt:PRL:2003,arndt:NatCommun:2011,arndt:RMP:2012,Kialka:AVSQuantSci:2022,Arndt:Nature:2026}. Standard open-system approaches based on master equations and reduced density matrices \cite{breuer-bk:2002} have provided a quantitative framework to describe the corresponding loss of spatial coherence in Talbot--Lau and Kapitza--Dirac--Talbot--Lau interferometers \cite{Hornberger:PRA:2004,joos-zeh:ZPhysB:1985,joos:bk:1996,schlosshauer:RMP:2005,schlosshauer-bk:2007}.

Most analyses of decoherence in interferometry focus on global observables, such as fringe visibility, intensity profiles, momentum distributions, and phase-space representations. These quantities are directly connected to measurable interference contrast and therefore provide the natural diagnostics of coherence loss. However, they do not necessarily provide a local description of how probability is transported across the interferometer. In particular, the disappearance of visible interference fringes does not by itself determine whether the underlying probability flow has lost all its spatial organization. This distinction is especially relevant in periodic near-field diffraction, where the grating geometry can impose a structured flow even when phase correlations between distant openings are progressively suppressed.

A useful way to address this local aspect is to analyze the probability current associated with the diffracted matter wave. The corresponding Bohmian, or hydrodynamic, representation is fully equivalent to the standard Schr\"odinger description and provides a direct visualization of local transport through the streamlines of the probability current
\cite{sanz:JCP-Talbot:2007,sanz:JPA:2008,sanz:AJP:2012,sanz:FrontPhys:2019}. In the context of grating diffraction, this representation has shown that Talbot carpets are accompanied by a characteristic channeling of the quantum flow, with spatial domains associated with the periodic arrangement of the slits \cite{sanz:JCP-Talbot:2007,sanz:AOP:2015}. Such a description does not introduce additional physical assumptions; rather, it reorganizes the standard quantum dynamics in terms of local flux and transverse momentum fields.

Trajectory-based analyses were also used previously to examine decoherence and contextuality in simpler interference arrangements. In particular, two-slit interference under environmental coupling was studied from a quantum-trajectory perspective in Refs.~\cite{sanz:EPJD:2007,sanz:CPL:2009-2}, where the loss of coherence was related to the progressive modification of the associated flow structure. More recently, decoherence-induced suppression of carpet-type structures was analyzed in confined geometries \cite{honrubia:PRA:2021}. The present work builds on this line of research but addresses a different AMO setting: matter-wave Talbot interference generated by a periodic grating. This geometry allows us to study how decoherence affects not only the visibility of near-field revivals but also the persistence and eventual degradation of flux channels induced by the grating periodicity.

Here we investigate decoherence in matter-wave Talbot interference using an effective open-system model combined with a probability-flow analysis. The incident beam is modeled as a monochromatic atomic matter wave diffracted by a periodic array of Gaussian apertures, a standard approximation for grating diffraction under paraxial conditions \cite{sanz:JPCM:2002,sanz:AOP:2015}. Environmental coupling is introduced through an exponential damping of spatial coherences in the reduced density matrix, following the usual phenomenology of collisional and localization-induced decoherence \cite{joos-zeh:ZPhysB:1985,joos:bk:1996,Hornberger:PRA:2004}. This model allows us to interpolate continuously between the fully coherent Talbot regime and the incoherent far-field diffraction limit while keeping the analysis analytically transparent.

The purpose of this paper is twofold. First, we characterize the suppression of Talbot carpets and the associated smoothing of transverse momentum distributions as the decoherence strength is increased. Second, we analyze the corresponding probability-flow streamlines in order to determine how the local channel structure evolves during the loss of spatial coherence. We show that, in this periodic matter-wave geometry, the reduction of visible multislit interference does not immediately entail the loss of the grating-defined organization of the local probability flow.
Thus, Talbot interferometry provides a useful platform to distinguish coherence loss at the level of phase correlations from the reorganization of probability transport.

This paper is organized as follows. In Sec.~\ref{sec2}, we introduce the paraxial description of matter-wave diffraction, the Gaussian-slit model for periodic gratings, and the effective decoherence model used throughout the work. We also summarize the probability-flow formulation and its relation to the transverse momentum field. In Sec.~\ref{sec3}, we present the numerical results for the coherent and decohering Talbot regimes, including intensity patterns, transverse momentum maps, and probability-flow streamlines. Finally, in Sec.~\ref{sec4}, we discuss the implications of the results for decoherence diagnostics in matter-wave interferometry and summarize the main conclusions.


\section{Theoretical framework}
\label{sec2}



\subsection{Paraxial matter-wave diffraction}
\label{sec21}

We consider a monochromatic beam of identical, noninteracting, spinless atoms of mass $m$ propagating predominantly along the $z$ direction. The incident beam is characterized by a de Broglie wavelength $\lambda_{\rm dB}$ and longitudinal wave number
\begin{equation}
    k = \frac{2\pi}{\lambda_{\rm dB}} ,
\end{equation}
so that the longitudinal momentum is $p_z=\hbar k$. Before diffraction, the matter wave can be represented, within the usual plane-wave approximation, as
\begin{equation}
    \Psi({\bf r},t)
    =
    \psi_0
    \exp\left[
        \frac{i}{\hbar}{\bf p}\cdot{\bf r}
        -
        \frac{i}{\hbar}Et
    \right],
    \label{planewave}
\end{equation}
with ${\bf p}=p_z\hat{\bf z}$ and $E=p_z^2/2m$. Since the longitudinal motion is essentially uniform, the propagation coordinate $z$ can be related to time through
\begin{equation}
    z = \frac{p_z}{m}t
      = \frac{\hbar k}{m}t .
    \label{ztrelation}
\end{equation}
This relation allows one to formulate the transverse dynamics either as a time-dependent one-dimensional problem or, equivalently, as a stationary propagation problem along $z$.

After the beam crosses the grating, the transverse momentum remains much smaller than the longitudinal one, $p_x\ll p_z$. Under this paraxial condition, the diffracted wave can be written as
\begin{equation}
    \Psi({\bf r},t)
    \simeq
    \psi(x,z)
    \exp\left[
        \frac{i}{\hbar}p_z z
        -
        \frac{i}{\hbar}Et
    \right],
\end{equation}
where $\psi(x,z)$ contains the transverse diffraction dynamics. Substitution into the free Schr\"odinger equation, neglecting second derivatives with respect to $z$ in comparison with transverse variations, leads to the paraxial matter-wave equation \cite{sanz:AOP:2015,sanz:ApplSci:2020}
\begin{equation}
    i\frac{\partial \psi(x,z)}{\partial z} =
    - \frac{1}{2k} \frac{\partial^2 \psi(x,z)}{\partial x^2}.
    \label{Helmholtz}
\end{equation}
Equation~(\ref{Helmholtz}) is formally equivalent to the one-dimensional free-particle Schr\"odinger equation after the identification~(\ref{ztrelation}). It is also the matter-wave analog of the paraxial Helmholtz equation used in scalar diffraction theory \cite{cronin:RMP:2009,sanz:ApplSci:2020,sanz:JOSAA:2012}. This equation will be the starting point for the propagation of all diffracted components.

\subsection{Probability flow and transverse momentum}
\label{sec22}

In addition to the intensity distribution,
\begin{equation}
    \rho(x,z)=|\psi(x,z)|^2 ,
\end{equation}
we analyze the local probability flow associated with Eq.~(\ref{Helmholtz}). Multiplying Eq.~(\ref{Helmholtz}) by $\psi^*$, subtracting the complex conjugate equation, and rearranging terms give the continuity equation
\begin{equation}
    \frac{\partial \rho(x,z)}{\partial z}
    = - \frac{\partial}{\partial x} \left[ \rho(x,z) v_{\rm eff}(x,z) \right],
    \label{continuity}
\end{equation}
where
\begin{equation}
\begin{aligned}
v_{\rm eff}(x,z) & = \frac{1}{2ik\rho(x,z)} \\ & \times
\left[\psi^*(x,z)\frac{\partial \psi(x,z)}{\partial x}
- \psi(x,z)\frac{\partial \psi^*(x,z)}{\partial x} \right].
\end{aligned}
\label{veloc}
\end{equation}
The quantity $v_{\rm eff}$ plays the role of a transverse drift field in the propagation variable $z$. It is not a velocity in the usual time-dependent sense, but it determines how the probability density is transported across the transverse coordinate as the beam propagates.

If the transverse wave function is written in polar form,
\begin{equation}
    \psi(x,z)=A(x,z)e^{iS(x,z)},
\end{equation}
with $A$ and $S$ being real, Eq.~(\ref{veloc}) becomes
\begin{equation}
    v_{\rm eff}(x,z)
    =
    \frac{1}{k}\frac{\partial S(x,z)}{\partial x}.
    \label{veffphase}
\end{equation}
Accordingly, the local transverse wave number is \cite{sanz:AOP:2015,sanz:ApplSci:2020}
\begin{equation}
    k_x(x,z)
    =
    k v_{\rm eff}(x,z)
    =
    \frac{\partial S(x,z)}{\partial x}.
    \label{transversmoment}
\end{equation}
The integral curves of the field $v_{\rm eff}$,
\begin{equation}
    \frac{dx}{dz}
    =
    v_{\rm eff}(x,z)
    =
    \frac{k_x(x,z)}{k},
    \label{streamline}
\end{equation}
define probability-flow streamlines.
In the language of the Bohmian representation \cite{sanz:FrontPhys:2019}, these streamlines are the corresponding trajectories; in the present work, however, they are used only as a representation of the standard probability current.
Although the streamlines are used here as a theoretical diagnostic, they are completely determined by the phase or, equivalently, by the probability current of the matter wave, and therefore encode information complementary to intensity and momentum distributions.
In optical two-slit interferometry, closely related average flow lines have been reconstructed using weak-measurement techniques \cite{kocsis:Science:2011}, which illustrates that such current-based representations can have operational meaning in interferometric contexts.
No additional dynamical postulates are introduced; Eqs.~(\ref{continuity})--(\ref{streamline}) are simply a hydrodynamic rewriting of the paraxial wave equation \cite{sanz:JCP-Talbot:2007,sanz:JPA:2008,sanz:AJP:2012,sanz:FrontPhys:2019}.

\subsection{Gaussian-slit model for a periodic grating}
\label{sec23}

We now focus the discussion on diffraction by a periodic grating of period $d$. The grating is assumed to be illuminated by a sufficiently broad incident beam so that $N$ openings contribute to the diffracted wave. Following the Gaussian-slit approximation \cite{feynman-bk1,sanz:JPCM:2002,sanz:JCP-Talbot:2007,sanz:AOP:2015}, each opening is represented by a Gaussian transmission function centered at $x_n$, with
\begin{equation}
    x_{n+1}-x_n=d .
\end{equation}
Immediately behind the grating, the transverse wave function is written as a coherent superposition of the partial amplitudes emerging from the illuminated openings,
\begin{equation}
    \psi(x,0)
    =
    \frac{\psi_{\rm out}}{\sqrt{N}}
    \sum_{n=1}^{N}
    \phi_n(x,0),
    \label{diffract0}
\end{equation}
where
\begin{equation}
    \phi_n(x,0)
    =
    \left(
        \frac{1}{2\pi\sigma_0^2}
    \right)^{1/4}
    \exp\left[
        -\frac{(x-x_n)^2}{4\sigma_0^2}
    \right].
    \label{Gausszero}
\end{equation}
The parameter $\sigma_0$ is the effective width of the diffracted beam associated with each slit. In the numerical examples below we take $\sigma_0=w/4$, where $w$ is the geometrical slit width. The normalization in Eq.~(\ref{diffract0}) was chosen with a $1/\sqrt{N}$ prefactor, consistent with the propagated field below and with the interpretation of $|\psi_{\rm out}|^2$ as the total diffracted intensity ($\sigma_0$ and $d$ are chosen here such that the overlap between adjacent Gaussians is negligible).

The substitution of
\begin{equation}
    \phi_n(x,z)
    =
    \frac{1}{\sqrt{2\pi}}
    \int
    \tilde{\phi}_n(k_x,z)e^{ik_xx}\,dk_x
    \label{fourier}
\end{equation}
into the paraxial propagation equation, Eq.~(\ref{Helmholtz}), yields
\begin{equation}
 \frac{\partial\tilde{\phi}_n(k_x,z)}{\partial z} = -\frac{i k_x^2}{2k}\tilde{\phi}_n(k_x,z),
\end{equation}
whose solution is the propagated spectral amplitude
\begin{equation}
    \tilde{\phi}_n(k_x,z)
    =
    \tilde{\phi}_n(k_x,0)
    \exp\left[
        -\frac{i k_x^2 z}{2k}
    \right].
    \label{tildephin}
\end{equation}
Now substituting Eq.~(\ref{tildephin}) into the inverse Fourier representation, Eq.~(\ref{fourier}), and expressing $\tilde{\phi}_n(k_x,0)$ in terms of $\phi_n(x',0)$ yield the Huygens--Fresnel propagation integral,
\begin{equation}
    \phi_n(x,z)
    =
    \sqrt{\frac{k}{2\pi i z}}
    \int
    \phi_n(x',0)
    \exp\left[
        \frac{ik(x-x')^2}{2z}
    \right]dx' .
    \label{huygensint}
\end{equation}
Substituting Eq.~(\ref{Gausszero}) into Eq.~(\ref{huygensint}) gives
\begin{equation}
    \phi_n(x,z)
    =
    \left(
        \frac{1}{2\pi\tilde{\sigma}_z^2}
    \right)^{1/4}
    \exp\left[
        -\frac{(x-x_n)^2}{4\sigma_0\tilde{\sigma}_z}
    \right],
    \label{phinz}
\end{equation}
where
\begin{equation}
    \tilde{\sigma}_z
    =
    \sigma_0
    \left(
        1+\frac{i z}{2k\sigma_0^2}
    \right) .
    \label{sigmatilde}
\end{equation}
From Eq.~\eqref{sigmatilde}, the real beam width reads as
\begin{equation}
    \sigma_z
    =
    \sigma_0
    \sqrt{
        1+
        \left(
            \frac{z}{2k\sigma_0^2}
        \right)^2
    } .
    \label{sigmaz}
\end{equation}
The total diffracted wave is therefore
\begin{equation}
    \psi(x,z)
    =
    \frac{\psi_{\rm out}}{\sqrt{N}}
    \left(
        \frac{1}{2\pi\tilde{\sigma}_z^2}
    \right)^{1/4}
    \sum_{n=1}^{N}
    \exp\left[
        -\frac{(x-x_n)^2}{4\sigma_0\tilde{\sigma}_z}
    \right].
    \label{wavez}
\end{equation}

From Eq.~(\ref{wavez}), the intensity distribution can be expressed as
\begin{equation}
\begin{aligned}
\rho_0(x,z) & = \frac{|\psi_{\rm out}|^2}{N} \sqrt{\frac{1}{2\pi\sigma_0\sigma_z}} \\
 & \qquad \times \sum_{n,n'=1}^{N} \cos\varphi_{nn'}(x,z) e^{-\beta_{nn'}(x,z)} ,
\label{rho0}
\end{aligned}
\end{equation}
with
%
%
\begin{equation}
\beta_{nn'}(x,z) := \frac{(x-x_n)^2+(x-x_{n'})^2}{4\sigma_z^2} ,
\end{equation}
and
\begin{equation}
\varphi_{nn'}(x,z) := \frac{z}{8k\sigma_0^2\sigma_z^2} \left[(x-x_n)^2-(x-x_{n'})^2\right].
\end{equation}
The diagonal contributions $n=n'$ describe the incoherent sum of the individual diffracted beams, whereas the terms with $n\neq n'$ contain the interference between different openings.

For a sufficiently extended periodic grating, the coherent superposition~(\ref{wavez}) gives rise to near-field self-imaging. The corresponding Talbot distance is
\begin{equation}
    z_T = \frac{2d^2}{\lambda_{\rm dB}} .
    \label{talbotdist}
\end{equation}
At integer multiples of $z_T$ the transverse intensity reproduces the grating periodicity, while fractional revivals occur at rational fractions of this distance. The resulting near-field pattern is the matter-wave Talbot carpet \cite{patorski:ProgOpt:1989,berry:JModOpt:1996,berry:PhysWorld:2001,xiao:AdvOptPhoton:2013,sanz:JCP-Talbot:2007,schleich:NJP:2013}.

Substituting Eq.~(\ref{wavez}) into Eq.~(\ref{veloc}) yields the corresponding probability-flow field. For the coherent grating, one obtains
\begin{widetext}
\begin{equation}
    v_{\rm eff}^{(0)}(x,z)
    =
    \frac{1}{4k^2\sigma_0^2\sigma_z^2}
    \frac{
    \displaystyle
    \sum_{n,n'=1}^{N}
    \left[
        z\cos\varphi_{nn'}(x,z)
        -
        2k\sigma_0^2\sin\varphi_{nn'}(x,z)
    \right]
    (x-x_n)
    e^{-\beta_{nn'}(x,z)}
    }{
    \displaystyle
    \sum_{n,n'=1}^{N}
    \cos\varphi_{nn'}(x,z)
    e^{-\beta_{nn'}(x,z)}
    } .
    \label{veloc0}
\end{equation}
\end{widetext}
This expression, derived and discussed in detail in previous analyses of Talbot diffraction and grating interferometry \cite{sanz:JCP-Talbot:2007,sanz:AOP:2015}, makes explicit the connection between interference phases, transverse momentum, and probability-flow channeling. For a periodic grating, the flow is organized into transverse domains associated with the grating openings. These domains underlie the channel structure observed in the Talbot carpet.

\subsection{Effective model of spatial decoherence}
\label{sec24}

To describe environmental loss of spatial coherence, we use an effective open-system model in which the off-diagonal elements of the reduced density matrix are exponentially damped in the position representation. The starting point is a Lindblad-type master equation of the form \cite{breuer-bk:2002}
\begin{equation}
    \frac{\partial \hat{\rho}(t)}{\partial t} =
    - \frac{i}{\hbar} [\hat{H},\hat{\rho}(t)]
    - \Gamma [\hat{X},[\hat{X},\hat{\rho}(t)]],
    \label{lindblad}
\end{equation}
where $\hat{\rho}$ is the reduced density matrix of the matter wave, $\hat{H}$ is the free-particle Hamiltonian, $\hat{X}$ is the transverse position operator, and $\Gamma$ characterizes the strength of the environmental localization mechanism. This type of position-localization term provides a standard phenomenological description of spatial decoherence induced by environmental scattering or monitoring \cite{joos-zeh:ZPhysB:1985,joos:bk:1996,schlosshauer:RMP:2005,schlosshauer-bk:2007,Hornberger:PRA:2004}.

For a free particle in one transverse dimension, Eq.~(\ref{lindblad}) becomes
\begin{widetext}
\begin{equation}
    \frac{\partial \rho(x,x',t)}{\partial t}
    =
    -
    \frac{i\hbar}{2m}
    \left(
        \frac{\partial^2}{\partial x^2}
        -
        \frac{\partial^2}{\partial x'^2}
    \right)
    \rho(x,x',t)
    -
    \Gamma (x-x')^2 \rho(x,x',t).
    \label{mastert}
\end{equation}
Using the relation between time and propagation distance [Eq.~(\ref{ztrelation})], we take the paraxial form of this equation as
\begin{equation}
    \frac{\partial \rho(x,x',z)}{\partial z}
    =
    -
    \frac{i}{2k}
    \left(
        \frac{\partial^2}{\partial x^2}
        -
        \frac{\partial^2}{\partial x'^2}
    \right)
    \rho(x,x',z)
    -
    \Lambda (x-x')^2 \rho(x,x',z),
    \label{masterz}
\end{equation}
\end{widetext}
with
\begin{equation}
    \Lambda
    =
    \frac{m\Gamma}{\hbar k} ,
    \label{Lambda}
\end{equation}
and only correlations along the transverse direction are taken into account.
When the decoherence term is used to describe the dominant suppression of spatial coherences, the off-diagonal damping is approximated by \cite{joos-zeh:ZPhysB:1985,joos:bk:1996}
\begin{equation}
    \rho(x,x',z)
    \simeq
    \rho(x,x',0)
    \exp[-\Lambda z (x-x')^2].
    \label{rhodecay}
\end{equation}
This expression leaves the diagonal density $\rho(x,x,z)$ unaffected while suppressing coherences between increasingly separated transverse positions.

In the grating problem, we implement this damping at the level of the interference terms between different Gaussian components. This provides a coarse-grained implementation of the position-space damping in Eq.~(\ref{rhodecay}), in which each diffracted Gaussian component is represented, for the purpose of estimating inter-slit decoherence, by the center of the corresponding grating opening. In this way, the single-slit diffraction dynamics is preserved, whereas the mutual coherence between distinct openings is progressively suppressed. Since the components associated with slits $n$ and $n'$ are centered at $x_n$ and $x_{n'}$, their mutual coherence is weighted by
\begin{equation}
    D_{nn'}(z) = \exp[-\Lambda z (x_n-x_{n'})^2].
    \label{Dnn}
\end{equation}
Accordingly, the decohering intensity distribution becomes
\begin{widetext}
\begin{equation}
    \rho(x,z)
    =
    \frac{|\psi_{\rm out}|^2}{N}
    \sqrt{
        \frac{1}{2\pi\sigma_0\sigma_z}
    }
    \sum_{n,n'=1}^{N}
    \cos\varphi_{nn'}(x,z)
    e^{-\beta_{nn'}(x,z)}
    e^{-\Lambda z(x_n-x_{n'})^2}.
    \label{rhodec}
\end{equation}
For $\Lambda=0$, Eq.~(\ref{rhodec}) reduces to the fully coherent Talbot intensity~(\ref{rho0}). In the opposite limit, the off-diagonal terms with $n\neq n'$ are suppressed, and the density approaches the incoherent sum of single-slit diffraction intensity distributions.

The corresponding probability-flow field is obtained by applying the same damping to the interference terms entering Eq.~(\ref{veloc0}). This gives
\begin{equation}
    v_{\rm eff}(x,z)
    =
    \frac{1}{4k^2\sigma_0^2\sigma_z^2}
    \frac{
    \displaystyle
    \sum_{n,n'=1}^{N}
    \left[
        z\cos\varphi_{nn'}(x,z)
        -
        2k\sigma_0^2\sin\varphi_{nn'}(x,z)
    \right]
    (x-x_n)
    e^{-\beta_{nn'}(x,z)}
    e^{-\Lambda z(x_n-x_{n'})^2}
    }{
    \displaystyle
    \sum_{n,n'=1}^{N}
    \cos\varphi_{nn'}(x,z)
    e^{-\beta_{nn'}(x,z)}
    e^{-\Lambda z(x_n-x_{n'})^2}
    } .
    \label{velocdec}
\end{equation}
\end{widetext}
Equations~(\ref{rhodec}) and~(\ref{velocdec}) are the central working expressions used in the numerical analysis. They describe, respectively, the suppression of Talbot interference in the intensity distribution and the modification of the local probability flow under the same decoherence mechanism. The construction is analogous to previous trajectory-based treatments of decoherence in interference and carpet-type structures \cite{sanz:EPJD:2007,sanz:CPL:2009-2,honrubia:PRA:2021} but is applied here to the periodic Talbot geometry relevant to matter-wave diffraction.

It is useful to associate Eq.~(\ref{Dnn}) with a propagation-dependent coherence range.
In this regard, if the damping between two components becomes of order $e^{-1}$ when
\begin{equation}
    \Lambda z (x_n-x_{n'})^2 \sim 1 ,
\end{equation}
then we define the coherence range as
\begin{equation}
 \Delta x := \frac{1}{\sqrt{\Lambda z}} ,
 \label{cohrange}
\end{equation}
which estimates the transverse separation over which mutual coherence remains appreciable at a given propagation distance. Accordingly, the ratio $\Delta x/d$ provides an estimate of the number of grating periods that remain mutually coherent.
This quantity will be used below to interpret the progressive disappearance of Talbot revivals and the persistence or degradation of probability-flow channels.


\begin{figure*}[!t]
    \centering
    \includegraphics[width=0.85\linewidth]{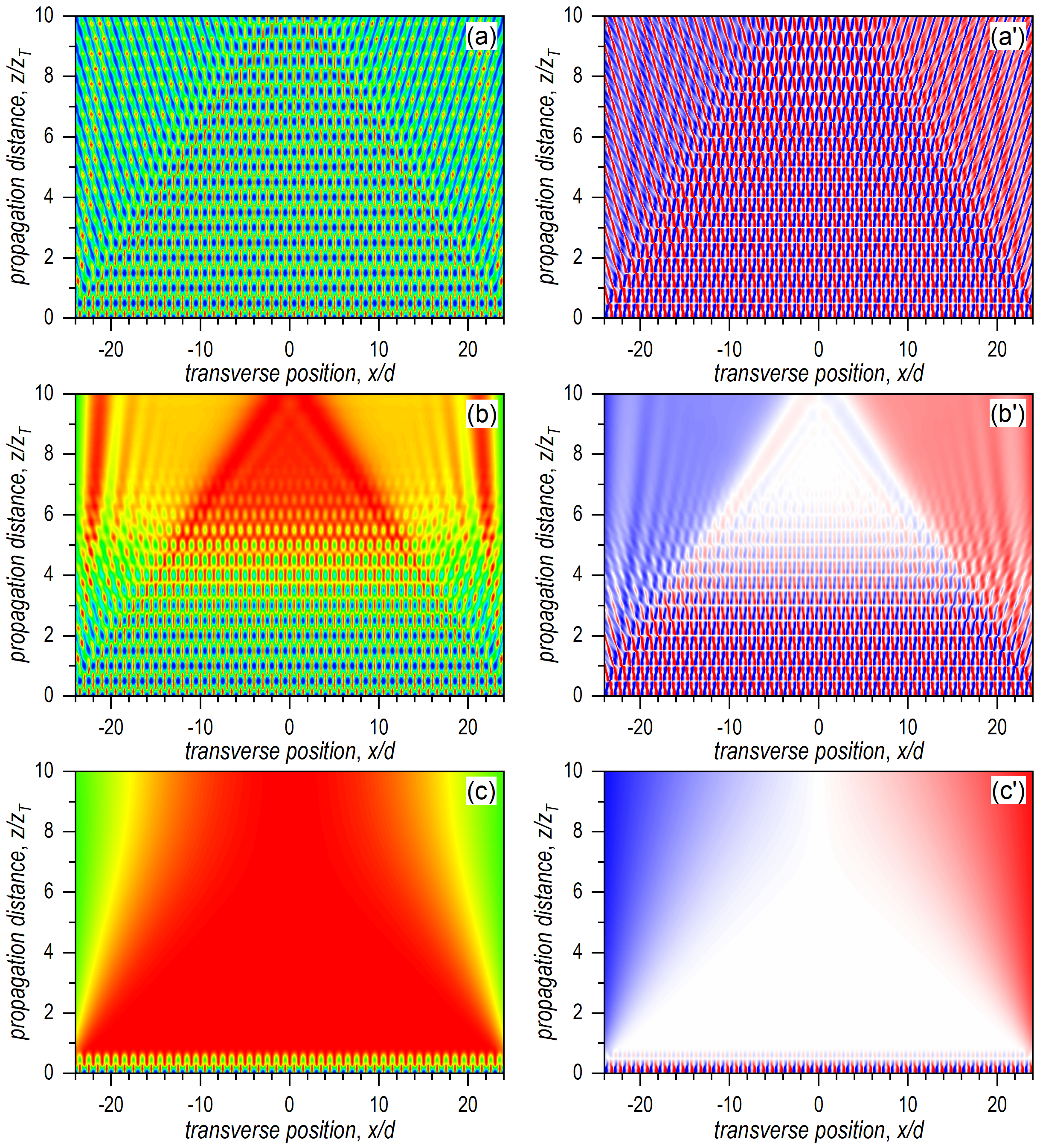}
    \caption{Talbot-like carpet produced by the superposition of a finite periodic array of 50 diffracted Gaussian wave packets and its gradual suppression as the decoherence strength is increased: (a) and (a') $\Lambda=0$, (b) and (b') $\Lambda=10^{-3}$~mm$^{-1}\!\cdot\!\mu$m$^{-2}$, and (c) and (c') $\Lambda=1$~mm$^{-1}\!\cdot\!\mu$m$^{-2}$. The density plots of the intensity are shown in the left column, while the right column shows the corresponding relative transverse momentum $k_x/k_0$, with $k_0=2\pi/d$. The blue-to-red color code in the left column denotes increasing normalized intensity.
 In the right column, the blue-to-red scale denotes negative-to-positive values of $k_x/k_0$ in the range $-0.5\le k_x/k_0\le 0.5$, with white corresponding to nearly zero transverse momentum.}
    \label{fig1}
\end{figure*}

\section{Results}
\label{sec3}

We now apply the model of Sec.~\ref{sec2} to a matter-wave grating configuration with parameters typical of atom-interferometry experiments. We consider a high flux of sodium atoms with de Broglie wavelength $\lambda_{\rm dB}=16$~pm ($k \simeq 3.93\times 10^{11}$~m$^{-1}$) impinging on a periodic grating of period $d=0.4$~$\mu$m and slit width $w=0.2$~$\mu$m \cite{keith:PRL:1991,chapman:PRL:1995}. The effective width of each diffracted Gaussian beam is taken as $\sigma_0=w/4=0.05$~$\mu$m, so that the intensity at the geometrical slit edges is already negligible. With these parameters, the Talbot distance \eqref{talbotdist} is $z_T=20$~mm, and the transverse grating wave number is
\begin{equation}
    k_0=\frac{2\pi}{d} \simeq 1.57\times 10^7~{\rm m}^{-1}.
\end{equation}
The simulations below are represented in terms of the dimensionless propagation and transverse coordinates introduced, namely $z/z_T$ and $x/d$. Unless otherwise stated, the grating is modeled by $N=50$ illuminated Gaussian openings. The quantities analyzed are the intensity distribution, Eq.~(\ref{rhodec}), the relative transverse momentum $k_x/k_0$, obtained from Eq.~(\ref{transversmoment}), and the probability-flow streamlines obtained by integrating Eq.~(\ref{streamline}).

\subsection{Near-field Talbot suppression}
\label{sec31}

\begin{figure*}[!t]
    \centering
    \includegraphics[width=0.85\linewidth]{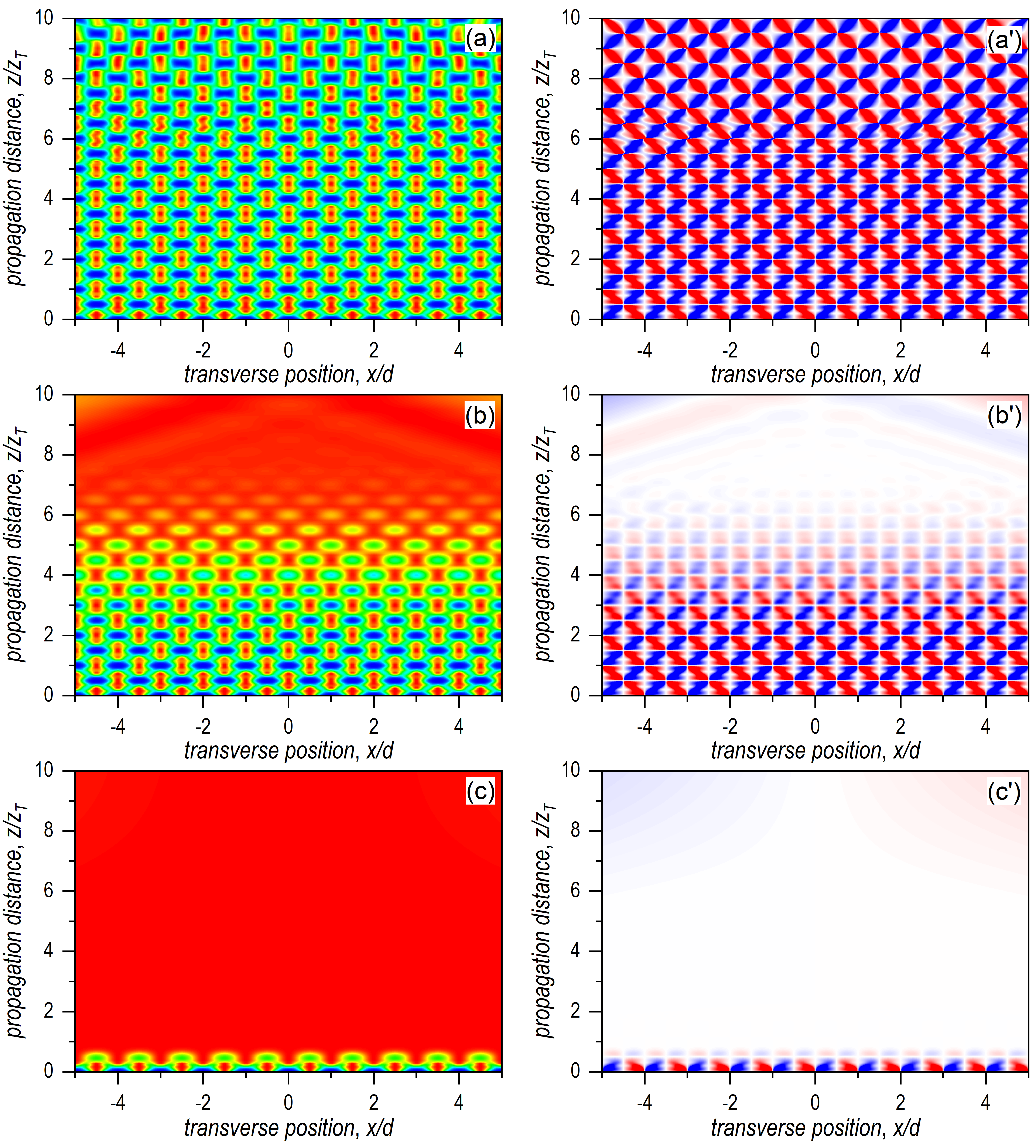}
    \caption{Same as Fig.~\ref{fig1}, but restricted to a neighborhood of $x=0$ in order to highlight the Talbot-like structure for $\Lambda=0$ and its progressive suppression as $\Lambda$ increases.}
    \label{fig2}
\end{figure*}

Figure~\ref{fig1}(a) shows the coherent Talbot carpet generated when $\Lambda=0$. The near-field self-imaging structure is clearly visible over many Talbot lengths, with the expected degradation at the outer boundaries due to the finite number of illuminated openings. This finite-size effect is caused by the absence of additional neighboring beams at the grating edges, which allows probability flux to leak outward. Nevertheless, around the central region the periodic Talbot structure remains robust, as shown in the magnified view of Fig.~\ref{fig2}(a).

The corresponding transverse momentum map, Fig.~\ref{fig1}(a'), displays a structured sequence of positive and negative momentum domains. This alternating pattern reflects the local modulation of the probability current induced by the grating periodicity. In particular, symmetry requires the transverse current to vanish at the midpoint between two symmetrically placed openings; consequently, the transverse momentum changes sign across such separatrices, in agreement with previous analyses of two-slit and grating diffraction \cite{sanz:JPA:2008,sanz:JCP-Talbot:2007}.

Figures~\ref{fig1}(b) and \ref{fig1}(c) show the effect of increasing environmental coupling. For weak decoherence, $\Lambda=10^{-3}$~mm$^{-1}\!\cdot\!\mu$m$^{-2}$, the Talbot structure survives for several Talbot lengths, although the contrast and fine structure are progressively reduced as propagation proceeds. For strong decoherence, $\Lambda=1$~mm$^{-1}\!\cdot\!\mu$m$^{-2}$, the self-imaging pattern is suppressed over much shorter distances, and the intensity evolves toward a smooth distribution dominated by the incoherent contribution of the individual diffracted beams. The corresponding transverse momentum maps, Figs.~\ref{fig1}(b') and \ref{fig1}(c'), show a progressive smoothing of the alternating momentum domains, especially around the central region. The enlarged panels in Fig.~\ref{fig2} make this transition more apparent.

\begin{figure}[!t]
 \centering
 \includegraphics[width=\columnwidth]{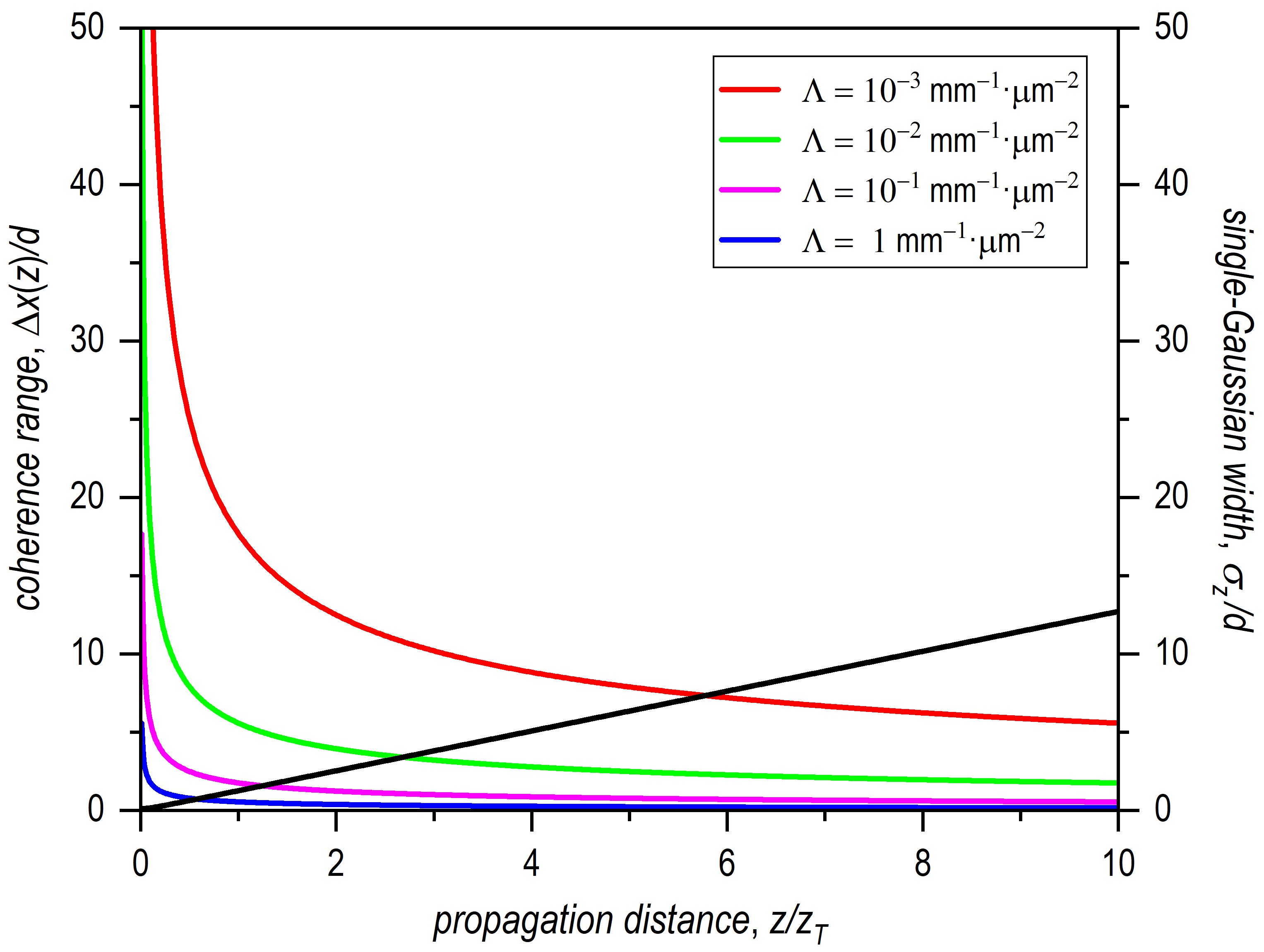}
 \caption{\label{fig3}
Dependence of the interslit coherence range $\Delta x$ on the propagation distance $z$ for several decoherence strengths: $10^{-3}$~mm$^{-1}\cdot\mu$m$^{-2}$ (red solid line), $10^{-2}$~mm$^{-1}\cdot\mu$m$^{-2}$ (green solid line), $10^{-1}$~mm$^{-1}\cdot\mu$m$^{-2}$ (magenta solid line), and $1$~mm$^{-1}\cdot\mu$m$^{-2}$ (blue solid line). The ratio $\Delta x/d$ estimates the number of grating periods over which mutual coherence remains appreciable. For comparison, the width $\sigma_z$ of a single diffracted Gaussian [Eq.~(\ref{sigmaz})] is also shown (black solid line; right axis).
}
\end{figure}

The suppression of the Talbot pattern can be interpreted in terms of the coherence range introduced in Eq.~(\ref{cohrange}). The quantity $\Delta x$ estimates the transverse separation over which mutual coherence between diffracted components remains appreciable at a given propagation distance $z$. Accordingly, the ratio $\Delta x/d$ provides an estimate of the number of grating periods that remain mutually coherent. In Fig.~\ref{fig3}, $\Delta x$ is plotted for several values of $\Lambda$ and compared with the width $\sigma_z$ of an individual diffracted Gaussian, Eq.~(\ref{sigmaz}).

As long as $\Delta x$ extends over several grating periods, different Gaussian components remain mutually coherent and Talbot-like structures can develop. When $\Delta x$ becomes comparable to or smaller than the transverse extent of an individual diffracted component, only short-range coherence remains and the Talbot carpet progressively disappears. For $\Lambda=10^{-3}$~mm$^{-1}\cdot\mu$m$^{-2}$, the coherence range becomes smaller than $\sigma_z$ at approximately $z\sim6z_T$, consistent with the degradation of the Talbot pattern observed in Figs.~\ref{fig1}(b) and \ref{fig2}(b). For the stronger decoherence rate $\Lambda=1$~mm$^{-1}\cdot\mu$m$^{-2}$, this crossover occurs before one Talbot distance is reached, in agreement with the rapid suppression shown in Figs.~\ref{fig1}(c) and \ref{fig2}(c). At larger propagation distances, the incoherent evolution of the broadened Gaussian components leads to the asymptotic single-envelope profile discussed below.

\subsection{Probability-flow streamlines in the near field}
\label{sec32}

\begin{figure*}[!t]
 \centering
 \includegraphics[width=0.85\linewidth]{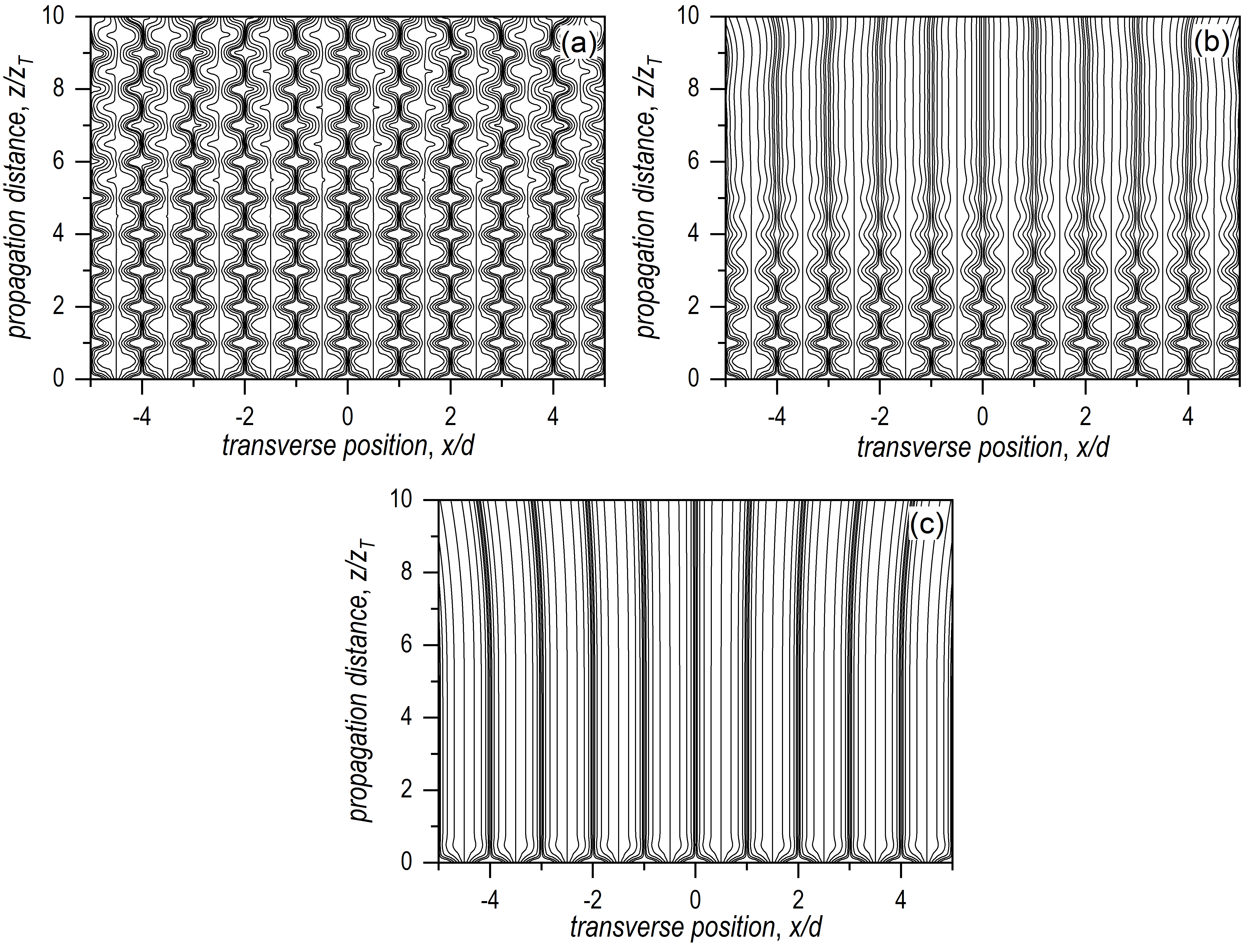}
 \caption{\label{fig4}
 Probability-flow streamlines associated with the three cases represented in Fig.~\ref{fig1}: (a) $\Lambda=0$, (b) $\Lambda=10^{-3}$~mm$^{-1}\!\cdot\!\mu$m$^{-2}$, and (c) $\Lambda=1$~mm$^{-1}\!\cdot\!\mu$m$^{-2}$. A set of 11 streamlines, with evenly distributed initial conditions across each slit width, is used for each opening, giving 550 streamlines in total. For clarity, the panels are restricted to the central region considered in Fig.~\ref{fig2}.
 Channel organization is identified here with the confinement of the streamline bundles within transverse domains associated with the original grating cells, rather than with the persistence of oscillatory recurrences alone.
}
\end{figure*}

The intensity and transverse-momentum maps provide global information about the loss of coherence, but they do not fully display the local organization of probability transport. This information is contained in the streamlines obtained from Eq.~(\ref{streamline}). Equivalently, these curves are the Bohmian trajectories associated with the same probability-current field. Figure~\ref{fig4} shows representative streamline sets for the same decoherence strengths considered in Fig.~\ref{fig1}.

In the coherent case, Fig.~\ref{fig4}(a), the streamlines exhibit recurrent oscillatory motion within well-defined transverse domains associated with the grating openings. These domains form the local channel structure underlying the Talbot carpet. Their boundaries are associated with approximately vanishing transverse current and are located near the separations between neighboring grating cells, as also indicated by the separatrices in the transverse-momentum maps. Because the grating is finite, the outermost channels are less constrained and gradually move outwards. This releases the effective transverse confinement exerted on the inner channels and eventually contributes to the disappearance of the near-field Talbot structure at large propagation distances.

To distinguish channel organization from the noncrossing property of the streamlines alone, we use here a local geometrical criterion. We regard a grating-induced channel as preserved as long as the streamline bundle originating from a given opening remains confined within a transverse domain associated with the corresponding grating cell, with neighboring bundles separated by boundaries of nearly vanishing transverse current. Channel persistence therefore refers to the preservation of the grating-defined partition of the local probability flow. It does not require the recurrent oscillatory motion or the high-contrast density revivals characteristic of coherent Talbot self-imaging.

When decoherence is introduced, Figs.~\ref{fig4}(b) and \ref{fig4}(c), the oscillatory recurrences characteristic of the coherent Talbot carpet are progressively suppressed. In the weak-decoherence case, Fig.~\ref{fig4}(b), the streamline motion loses much of its recurrent structure, but the bundles remain organized within transverse domains associated with the original grating cells. This organization persists over propagation regions where the corresponding density patterns in Figs.~\ref{fig1}(b) and \ref{fig2}(b) have already undergone a substantial reduction of their Talbot structure. In this sense, the reduction of inter-slit coherence described by the coherence range in Fig.~\ref{fig3} does not immediately entail the loss of the grating-defined organization of the local probability flow.

For the stronger decoherence rate, Fig.~\ref{fig4}(c), both the recurrent motion and the grating-defined flow-domain structure are rapidly degraded. The streamlines increasingly approach the behavior expected from an incoherent set of independently diffracting Gaussian components. The comparison between Figs.~\ref{fig3} and \ref{fig4} therefore indicates two related but distinct changes: the suppression of the phase coherence required for Talbot self-imaging and the distinct degradation of the local flow organization inherited from the periodic grating.

This behavior is consistent with the construction of the model. The damping factor in Eq.~(\ref{Dnn}) suppresses phase coherence between different grating openings, but it does not alter the original aperture geometry or the individual single-slit contributions. The grating can consequently continue to organize the local probability transport after its high-contrast interference pattern has been substantially reduced. The streamline analysis thus complements the usual intensity-based description by separating the loss of Talbot self-imaging from the degradation of the grating-induced flow domains.

\begin{figure*}[!t]
    \centering
    \includegraphics[width=0.85\linewidth]{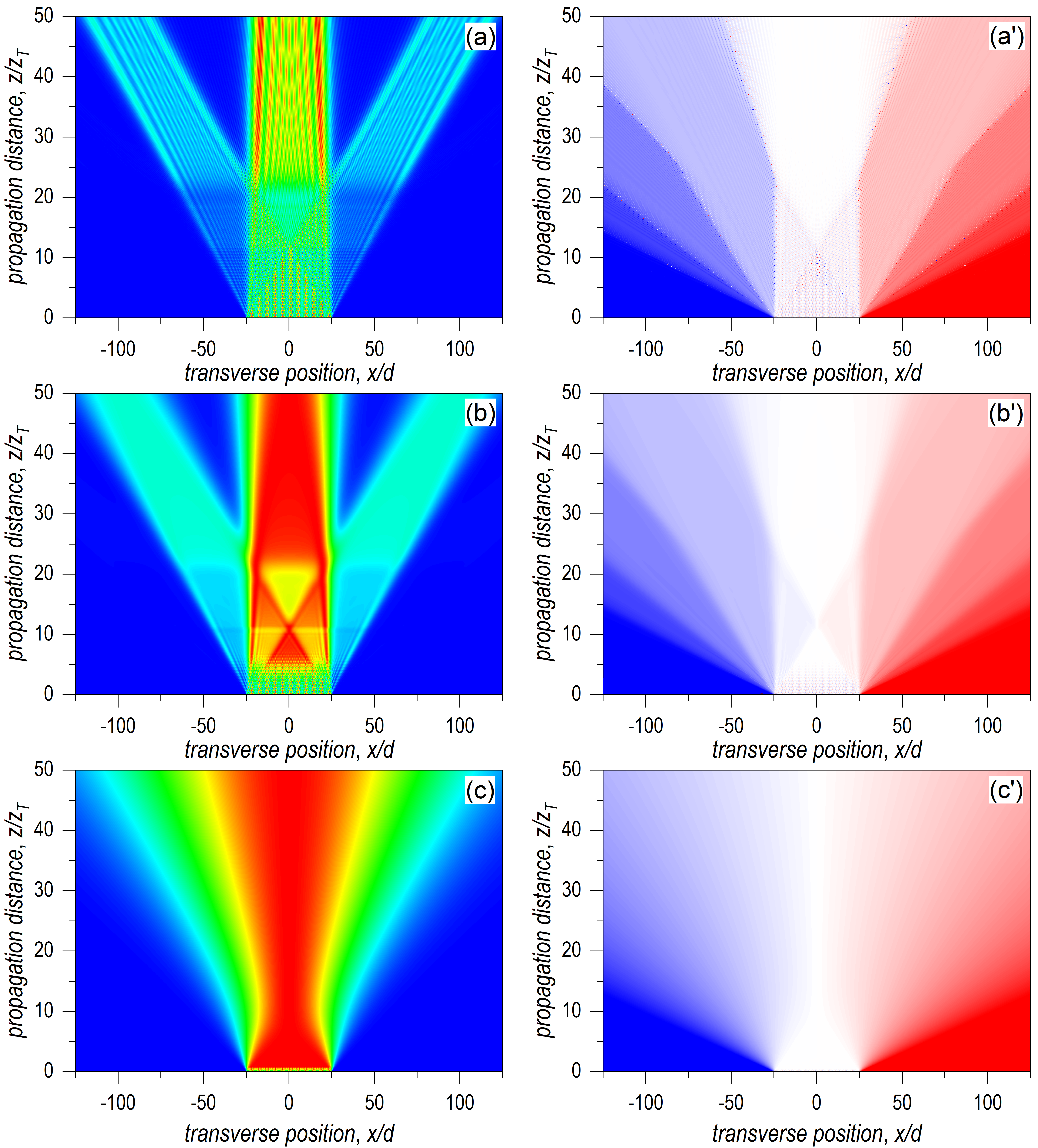}
    \caption{Far-field diffraction structures produced by the superposition of 50 diffracted Gaussian wave packets and their gradual suppression as the decoherence strength is increased: (a) and (a') $\Lambda=0$, (b) and (b') $\Lambda=10^{-3}$~mm$^{-1}\!\cdot\!\mu$m$^{-2}$, and (c) and (c') $\Lambda=1$~mm$^{-1}\!\cdot\!\mu$m$^{-2}$. The left column shows the normalized intensity, while the right column shows the relative transverse momentum $k_x/k_0$. In the right column, the color scale spans $-4\le k_x/k_0\le 4$, with white corresponding to nearly zero transverse momentum.}
    \label{fig5}
\end{figure*}

\subsection{Far-field transition and asymptotic profiles}
\label{sec33}

We next inspect the propagation at larger distances, where the system crosses over from the near-field Talbot regime to the far-field diffraction regime. Figure~\ref{fig5} shows the intensity and relative transverse momentum distributions from $z=0$ to a final propagation distance of $z_f=1$~m, corresponding to $50z_T$ for the present parameters. In the coherent case, Fig.~\ref{fig5}(a), a Fraunhofer-like diffraction structure is obtained. The intensity is concentrated in well-defined diffraction channels, with a dominant central order and weaker lateral orders. In the corresponding momentum map, Fig.~\ref{fig5}(a'), each channel is associated with an approximately quantized transverse momentum. The main regions correspond to $k_x/k_0\simeq 0,\pm 1,\pm 2,\ldots$, as expected for diffraction by a periodic grating.

For weak decoherence, Fig.~\ref{fig5}(b), the diffraction orders remain visible but the fine oscillatory structures within each channel are smoothed out. The momentum map, Fig.~\ref{fig5}(b'), still shows step-like regions associated with the main diffraction orders, although the transitions between adjacent values become less abrupt. Thus, even at propagation distances much larger than the Talbot length, partial coherence across the grating can leave signatures in both the intensity and transverse momentum distributions.

For strong decoherence, Fig.~\ref{fig5}(c), the far-field pattern loses the multi-order grating structure and approaches the broadened profile associated with incoherent single-slit diffraction. The corresponding transverse momentum distribution becomes approximately linear in the transverse coordinate, as expected for the free spreading of an individual Gaussian wave packet. In this regime, the information associated with the grating periodicity is largely washed out from the global observables.

\begin{figure*}[!t]
    \centering
    \includegraphics[width=0.85\linewidth]{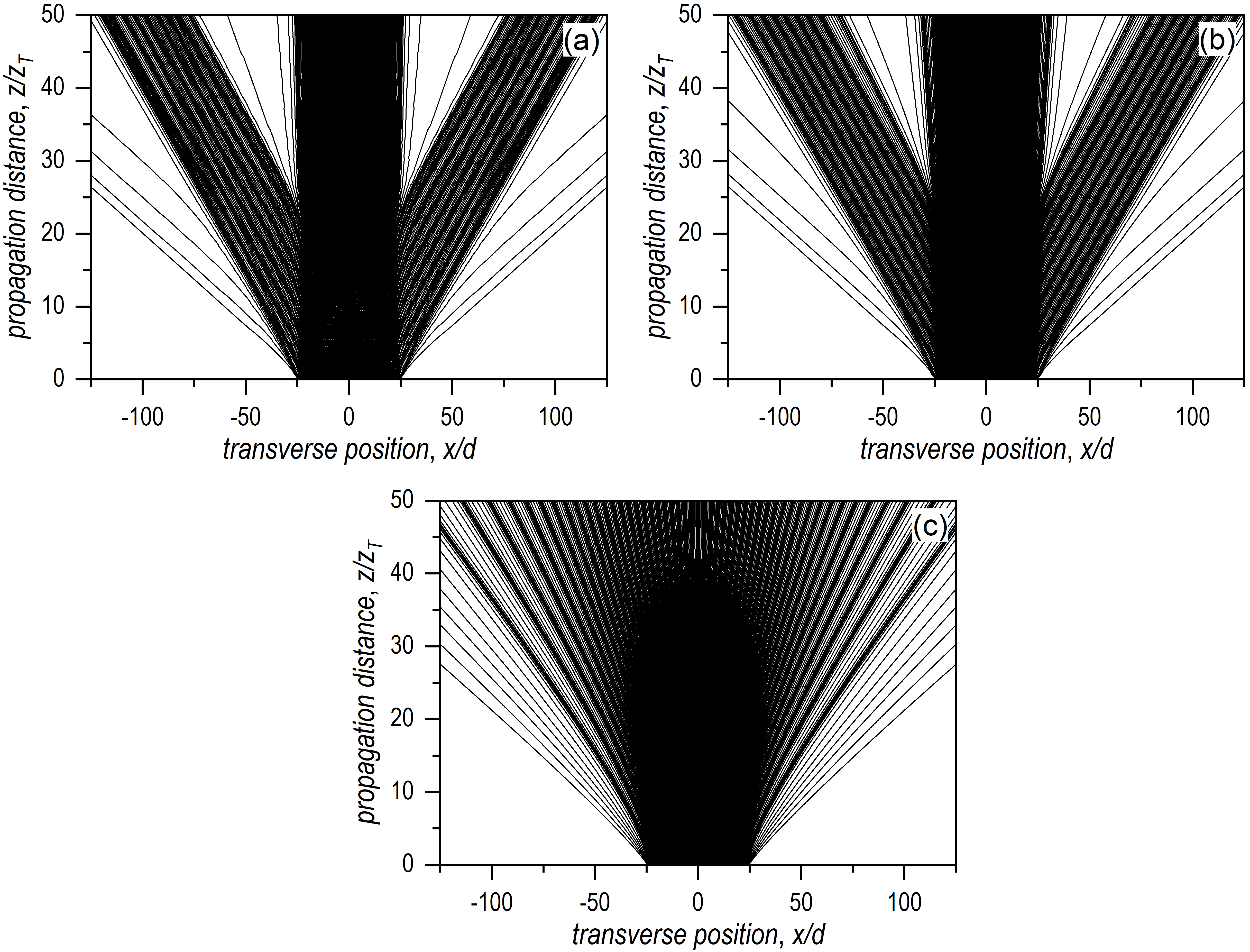}
    \caption{Probability-flow streamlines associated with the far-field cases represented in Fig.~\ref{fig5}: (a) $\Lambda=0$, (b) $\Lambda=10^{-3}$~mm$^{-1}\!\cdot\!\mu$m$^{-2}$, and (c) $\Lambda=1$~mm$^{-1}\!\cdot\!\mu$m$^{-2}$. All 550 streamlines are shown. For $\Lambda=0$ and weak decoherence, the density of streamlines highlights the main diffraction channels. For strong decoherence, the streamlines spread more homogeneously across the transverse plane, reflecting the suppression of the grating-induced channel structure.}
    \label{fig6}
\end{figure*}

The streamline representation of the same far-field regimes is shown in Fig.~\ref{fig6}. In the coherent case, Fig.~\ref{fig6}(a), the initially channelized near-field flow evolves into well-separated far-field diffraction beams. The central beam contains the largest fraction of the streamlines, in agreement with the dominant central diffraction order in Fig.~\ref{fig5}(a). For weak decoherence, Fig.~\ref{fig6}(b), the main pathways remain identifiable, although the internal fine structure is reduced. For strong decoherence, Fig.~\ref{fig6}(c), the streamline ensemble spreads over the transverse plane without forming sharply defined diffraction channels. This behavior mirrors the transition observed in the intensity and transverse momentum maps.

\begin{figure*}[!t]
    \centering
    \includegraphics[width=0.85\linewidth]{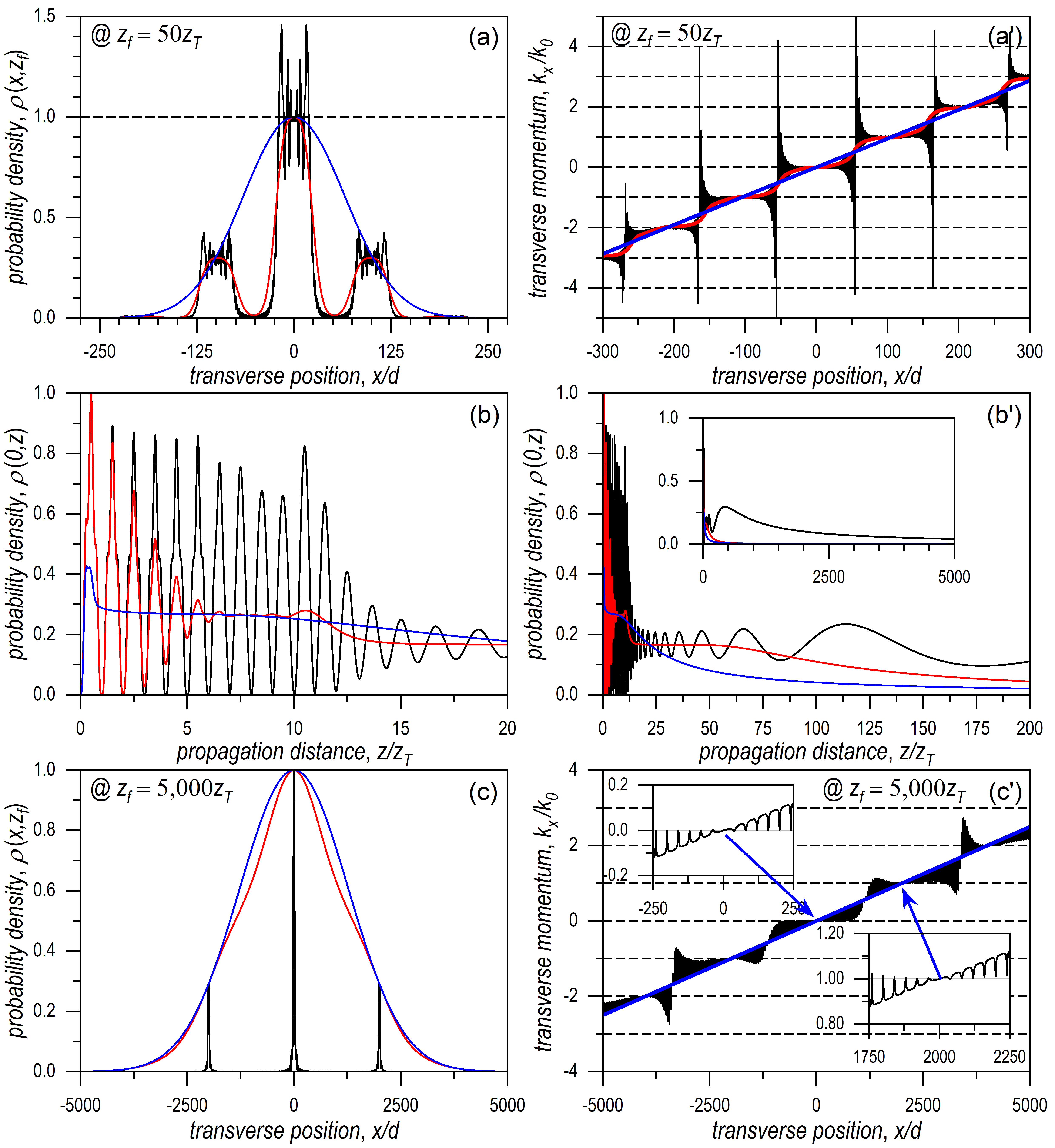}
    \caption{Asymptotic profiles of the intensity and transverse momentum for different decoherence strengths. (a) and (a') show, respectively, the probability density and relative transverse momentum at $z_f=50z_T$ for $\Lambda=0$ (black solid line), $\Lambda=10^{-3}$~mm$^{-1}\!\cdot\!\mu$m$^{-2}$ (red solid line), and $\Lambda=1$~mm$^{-1}\!\cdot\!\mu$m$^{-2}$ (blue solid line).
    As a reference, in (a) the horizontal dashed line denotes the average maximum (normalized to unity) of the 50-Gaussian wave-packet superposition.
  (b) and (b') show the on-axis probability density as a function of $z$, within and beyond the Talbot region, respectively. The inset in (b') displays the far-field decay.
  (c) and (c') show the same quantities as panels (a) and (a'), but at $z_f=5000z_T$.
    In (a') and (c'), the horizontal dashed lines denote the quantized values of the transverse momentum $k_x/k_0$, i.e., the so-called diffraction orders.}
    \label{fig7}
\end{figure*}

A more quantitative view of the coherent-to-incoherent crossover is provided in Fig.~\ref{fig7}. At $z_f=50z_T$, the coherent profile displays the multi-order structure associated with grating diffraction. Weak decoherence reduces the fine modulation and smooths the diffraction maxima, while strong decoherence removes the multislit structure and yields a single broad envelope. The transverse momentum profiles show the same trend: the coherent case exhibits sharp step-like transitions between diffraction orders, weak decoherence rounds these transitions, and strong decoherence produces an almost linear dependence characteristic of a freely spreading Gaussian beam.

The far-field positions of the coherent diffraction orders are consistent with the grating relation
\begin{equation}
    x_\ell \simeq \ell\frac{\lambda_{\rm dB}z}{d},
    \qquad
    \ell=0,\pm1,\pm2,\ldots .
    \label{difforderpos}
\end{equation}
For the parameters considered here and $z=z_f=1$~m, this gives $x_\ell\simeq 40\ell$~$\mu$m. The separation between adjacent momentum steps in Fig.~\ref{fig7}(a') is consistent with this order spacing. In the fully coherent limit, the far-field intensity may be compared with the usual grating expression
\begin{equation}
    \frac{\rho(x,z)}{\rho(x,0)}
    =
    e^{-x^2/2\sigma_z^2}
    \left[
        \frac{\sin(\pi dN\sin\theta/\lambda_{\rm dB})}
             {\sin(\pi d\sin\theta/\lambda_{\rm dB})}
    \right]^2,
    \label{multislit}
\end{equation}
which, under paraxial conditions and for large $z$, reduces to
\begin{equation}
    \frac{\rho(x,z)}{\rho(x,0)}
    \approx
    e^{-2k^2\sigma_0^2(x/z)^2}
    \left[
        \frac{\sin(\pi dNx/\lambda_{\rm dB}z)}
             {\sin(\pi dx/\lambda_{\rm dB}z)}
    \right]^2 .
    \label{multislit2}
\end{equation}
The centers of the diffraction maxima follow Eq.~(\ref{difforderpos}), while the detailed peak shapes are modified by the finite number of Gaussian openings and by the near-to-far-field crossover encoded in the propagated Gaussian envelope.

At very large distances, Fig.~\ref{fig7}(c), the differences between the coherent and partially decohered cases are further reduced in the intensity profiles, whereas the transverse momentum retains a clear distinction between step-like grating diffraction and the linear trend associated with incoherent Gaussian spreading. The on-axis profiles in Figs.~\ref{fig7}(b) and \ref{fig7}(b') show the gradual decay of the Talbot recurrences and the approach to the expected far-field decrease of the propagated density.

Taken together, the results show that the effective decoherence model produces a continuous crossover from coherent Talbot self-imaging to incoherent diffraction. At the same time, the probability-flow analysis indicates that the disappearance of visible Talbot fringes does not immediately entail the loss of the grating-induced organization of the local probability flow.
This distinction is the main dynamical point of the present analysis.


\section{Discussion and conclusions}
\label{sec4}

We analyzed the effect of spatial decoherence on matter-wave Talbot interference using an effective open-system model combined with a local probability-flow description. The system considered here is a periodic grating illuminated by an atomic beam under paraxial conditions, for which the coherent dynamics gives rise to the usual near-field Talbot carpet and, at longer propagation distances, to Fraunhofer-like diffraction orders. Decoherence was introduced through an exponential damping of the spatial coherences between diffracted components, which provides a simple and analytically transparent way to follow the crossover from coherent multislit diffraction to the incoherent addition of single-slit contributions.

A useful feature of the grating geometry is that the same framework follows the system from near-field Talbot self-imaging to far-field diffraction orders. This makes it possible to distinguish which features of the transport are tied to coherent self-imaging and which reflect the geometrical organization imposed by the initial periodic aperture structure.

At the level of standard observables, the model reproduces the expected behavior. As the decoherence strength increases, the Talbot recurrences lose contrast, the fine structure of the near-field carpet is progressively washed out, and the transverse momentum distribution becomes smoother. In the far field, the step-like momentum structure associated with grating diffraction is rounded for weak decoherence and evolves toward the approximately linear transverse-momentum profile characteristic of a freely spreading Gaussian beam for strong decoherence. The coherence-range estimate $\Delta x$ provides a useful interpretation of this behavior: Talbot features remain visible as long as several neighboring grating openings maintain appreciable mutual coherence, whereas the pattern rapidly disappears once the effective coherence range becomes comparable to the transverse extent of an individual diffracted component.

The probability-flow analysis provides complementary local information. In the coherent Talbot regime, the current streamlines display recurrent oscillatory motion within transverse domains associated with the grating openings. Under decoherence, the oscillatory recurrences and the high-contrast density revivals are progressively suppressed, but the grating-defined partition of the flow can remain recognizable over a longer propagation range. As specified in Sec.~\ref{sec32}, channel persistence refers here to the confinement of the streamline bundles within transverse domains associated with the original grating cells, with neighboring bundles separated by boundaries of approximately vanishing transverse current. It does not refer merely to the noncrossing property of the streamlines, nor does it require the persistence of Talbot self-imaging.

This distinction should not be interpreted as evidence of residual long-range phase coherence. The damping factor used here suppresses phase correlations between spatially separated grating openings and therefore reduces the interference terms responsible for the Talbot carpet. However, it does not modify the original aperture geometry or suppress the individual single-slit contributions. The periodic grating can consequently leave a geometrical imprint on the local probability-current field after its high-contrast interference pattern has been substantially reduced. The persistence of channeling is therefore understood as a local geometrical remnant of the initial periodic structure rather than as a surviving coherent interference effect.

The additional physical insight provided by this distinction is that loss of interference visibility does not constitute a complete description of how matter-wave transport is reorganized under decoherence. Visibility and density profiles characterize the suppression of phase-coherent interference, whereas the probability current describes the local direction and redistribution of the transmitted flux. A reduced Talbot contrast can therefore coexist with a current field that still retains a spatial organization related to the grating. This information is relevant for the interpretation of matter-wave transport whenever a structured flux distribution persists after the fine interferential features have become difficult to resolve.

The streamlines are not introduced here as an additional independent observable. They are fully determined by the probability current, or equivalently by the phase gradient and density of the matter wave. Their role is diagnostic: they provide a spatially resolved representation of information that remains implicit when only the intensity or its visibility is considered. In this sense, the hydrodynamic analysis complements, rather than replaces, standard intensity- and momentum-based observables. A direct experimental implementation of this diagnostic would require access to the local phase or current field and lies beyond the scope of the present work. Nevertheless, the analysis identifies which local transport features would need to be reconstructed in order to distinguish the loss of coherent self-imaging from the degradation of the grating-defined flow organization.

The present work extends previous trajectory-based studies of decoherence in simpler interference arrangements and confined carpet-like systems to the AMO-relevant setting of matter-wave Talbot diffraction by a periodic grating.
Talbot interferometry is particularly useful in this context because it combines the near-field recurrence of a periodic wave front with the coherent build up of multiple far-field diffraction orders. The present description follows both regimes within the same model and shows how spatial-coherence damping affects the density, transverse momentum, and local probability current.

The effective decoherence model used here is intentionally minimal. It captures the essential phenomenology of spatial localization and provides closed expressions for the decohering intensity and probability-flow field. This simplicity isolates the role of spatial-coherence damping without introducing environment-specific parameters that would obscure the connection between coherence loss and flow reorganization. At the same time, it defines the limitations of the analysis. More microscopic descriptions could incorporate the properties of the environmental scatterers, recoil distributions, non-Markovian effects, or time-dependent coupling strengths. Such extensions would be required for a quantitative comparison with a specific decohering environment.
The aim here is instead to isolate the dynamical consequences of spatial-coherence damping in a controlled Talbot geometry.

In summary, matter-wave Talbot interference provides a sensitive platform for examining decoherence in periodic near-field diffraction.
Environmental damping continuously transforms the coherent Talbot carpet into an incoherent diffraction pattern. During this crossover, the recurrent interferential structure and the grating-defined organization of the local probability flow do not necessarily disappear simultaneously. By making this distinction explicit, the hydrodynamic representation complements standard intensity and transverse-momentum diagnostics and provides a local characterization of the reorganization of matter-wave transport under decoherence.


\section*{Acknowledgments}


Research funded by MCIN/AEI/10.13039/501100011033/ and by FEDER, UE (PID2021-127781NB-I00).


\section*{Data availability}

The data that support the findings of this article are not publicly available.
The data are available from the authors upon reasonable request.



%

\end{document}